\documentclass[aps,pre,reprint,nobalancelastpage]{revtex4-1}

\usepackage[english]{babel}
\usepackage[latin1]{inputenc}

\usepackage{graphics}
\usepackage[dvipsnames]{graphicx}
\usepackage{xcolor}

\usepackage{amsfonts}
\usepackage{amssymb}

\usepackage[normalem]{ulem}

\definecolor{Pergamen}{RGB}{235,225,200}
\definecolor{LightGray}{RGB}{235,235,230}
\definecolor{Blue}{RGB}{0,0,255}

\definecolor{Red}{RGB}{255,0,0}

\usepackage{amsmath}

\begin{document}
\title{
	{\bf Cell-size distribution and scaling in a one-dimensional KJMA lattice model with continuous nucleation}
}


\author{Zolt\'an~N\'eda\thanks{zneda@ubbcluj.ro}, Ferenc J\'arai-Szab\'o and Szil\'ard Boda }
\affiliation{%
	Babe\c{s}-Bolyai University, Department of Physics, Cluj, Romania
}

\date{\today}


\newcommand{\vs}{\vspace{3mm}}

\newcommand{\be}{\begin{equation}}
\newcommand{\ee}[1]{\label{#1} \end{equation}}
\newcommand{\ba}{\begin{eqnarray}}
\newcommand{\ea}[1]{\label{#1} \end{eqnarray}}
\newcommand{\nl}{\nonumber \\}
\newcommand{\re}[1]{(\ref{#1})}
\newcommand{\spr}[2]{\vec{#1}\cdot\vec{#2}}
\newcommand{\ave}{\overline{u}}
\newcommand{\ve}[1]{\left\vert #1  \right\vert}
\newcommand{\exv}[1]{ \left\langle {#1} \right\rangle}

\newcommand{\pd}[2]{ \frac{\partial #1}{\partial #2}}
\newcommand{\pt}[2]{ \frac{{\rm d} #1}{{\rm d} #2}}
\newcommand{\pv}[2]{ \frac{\delta #1}{\delta #2}}

\newcommand{\grad}{{\vec{\nabla}}}

\newcommand{\ead}[1]{ {\rm e}^{#1}}
\newcommand{\infi}{ \int_0^{\infty}\limits\!}
\newcommand{\sumi}{ \sum_{n=0}^{\infty}\limits\!}


\begin{abstract}

The Kolmogrov-Johnson-Mehl-Avrami (KJMA) growth model is considered on a one-dimensional (1D) lattice. 
Cells can growth with constant speed and continuously nucleate on the empty sites.  We offer an alternative,
mean-field like approach for describing theoretically the dynamics and derive an analytical cell-size distribution function. 
Our method reproduces the same scaling laws as the KJMA theory and has the advantage that it leads to a simple closed form for the
cell-size distribution function. It is shown that a Weibull distribution is appropriate for describing the final cell-size distribution.
The results are discussed in comparison with Monte Carlo simulation data. 

\end{abstract}

\maketitle

\section{Introduction}

The  Kolmogrov-Johnson-Mehl-Avrami (KJMA) growth model \cite{Kolmogorov,Johnson,Avrami} has large applicability for describing several natural phenomena \cite{Fanfoni,Evans,Ramos} like domain growth associated with isothermal phase transformation \cite{Christian,Ishibashi,Henderson,Hirsch}, random sequential adsorption processes \cite{Feder,Evans} and thin film growth \cite{Lewis}. Recently, the model found exotic applications in molecular biology \cite{Herrick} and cosmology \cite{Kampfer}, as well.  The Voronoi-type space tessellations \cite{Voronoi} that appear as result of growth have also diverse scientific applications in physics, biology, material science, computer science, astrophysics, medicine, economics and sociology (see for example \cite{Voronoi,Jarai,Tomellini} and the references within these works).

Although the KJMA model has been extensively studied (see for example \cite{Axe,Sekimoto,Mulheran,Crespo,Jun,Farjas}) and in one-dimension (1D)  exact results are known for the cell size distribution function \cite{Axe,Ben-naim,Farjas}, the non-analytical form of the results limit their applicability.  We propose here an alternative, approximative theory, different from the one known to the KJMA process. Due to the involved approximations our results are not exact and less accurate than the one given by the the KJMA theory in 1D. However, the advantage of our approach is that it leads to the same scaling laws as the exact theory in 1D and suggests a simple analytical form for the cell-size distribution function. The situation is somehow similar with the case of Poissonian Voronoi cells size distribution function, where a simple and reasonably fair fit with a few parameters proves to be more useful for practical applications than a complicated, albeit more exact form \cite{Jarai}.    

In the followings first we present our lattice model and adapt it's results for the 1D KJMA process. Than we present our simple mean-field like 
analytical approximation for the studied quantities. Finally we present Monte Carlo type computer simulation data for this growth process and compare critically the obtained results with the theoretical ones.  

\section{The lattice model}

\begin{figure}[h]
\centerline{\includegraphics[width=0.45\textwidth]{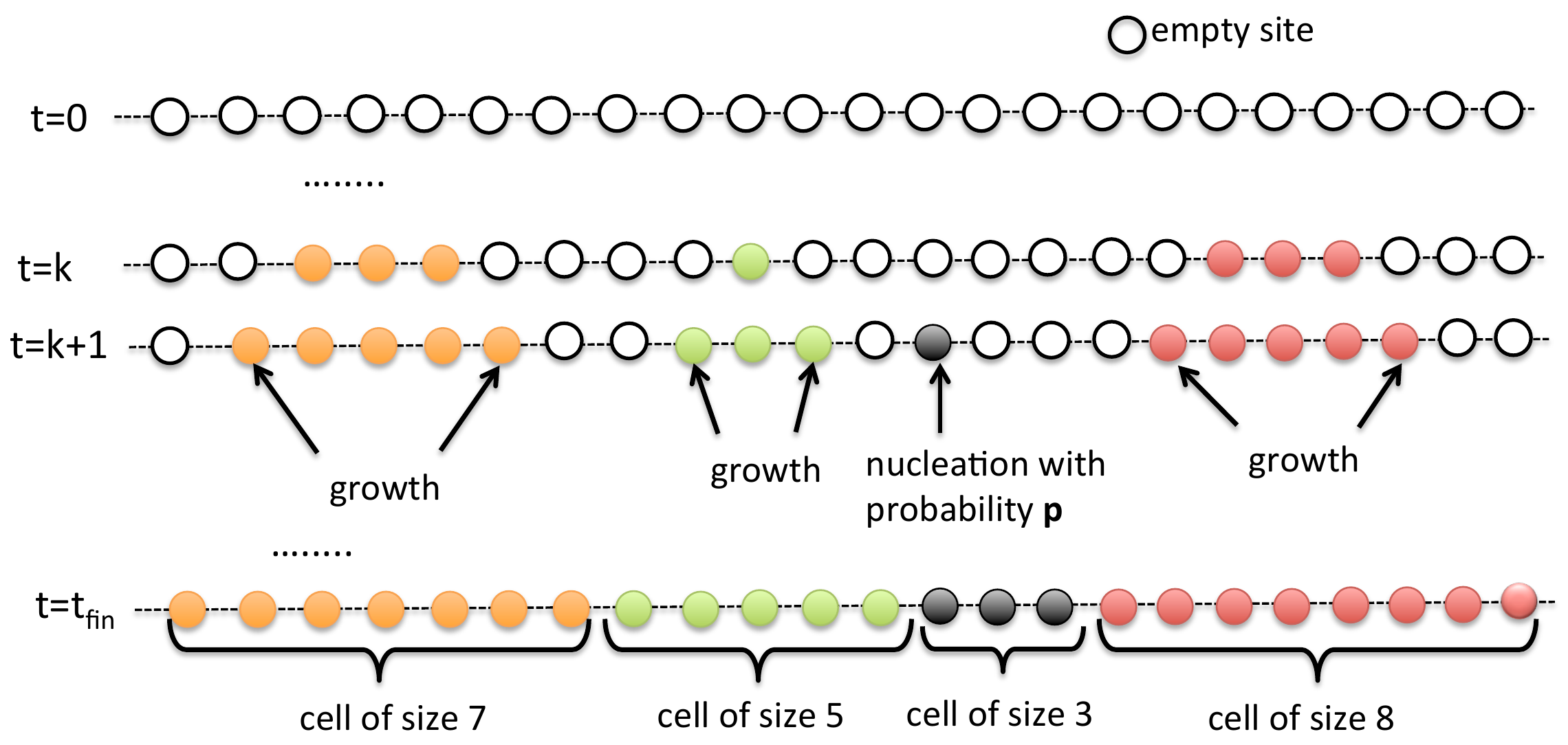} }
\caption{~\label{model}
Sketch of the 1D KJMA lattice model dynamics. } 
\end{figure}

The model considered here is the lattice version of the KJMA model in 1D. The growth process is sketched in Figure \ref{model}. At $t=0$ time-step all sites are empty. At any $t>0$ time-step, each empty site can be activated with a probability $p$, leading to a new Voronoi cell (depicted in the figure with different colours). The existing Voronoi cells will grow at both of their endpoints by occupying the neighbouring empty cells. Whenever two Voronoi cells get in contact, the growth stops at that boundary. The above nucleation and growth dynamics continues until all sites will be occupied by these Voronoi-like cells. The above described stochastic growth model has two parameters: the size of the one-dimensional lattice, $L$, and the nucleation probability, $p$.    

We are investigating the time $t_{fin}$, needed to get saturation and the statistics of the Voronoi-like cells at the end of the growth process. More specifically, we would like to determine the mean cell-size $c$, and the probability distribution function for the cell sizes. Our aim is to get a compact analytical approximation for the distribution function, one that could be useful in practical applications. 

\section{The classical KJMA theory in 1D}

The KJMA model corresponding to our growth model is the one which is known in the literature as the {\em continuous nucleation model} \cite{Kolmogorov,Johnson,Avrami}, where each cell (grain) is physically distinguishable from the others (or represent a different phase). The model is defined in continuous space and time. In  our discrete growth mechanism we define the length unit $l$, as the distance between two neighbouring lattice sites, and the unit time $T$ is set to one growth step, as it is described earlier. The original model in its general form is governed by two parameters, $p$ the new cell nucleation rate per unit length and $v$, the linear velocity at which the cells grow in both direction until impeded by a neighbouring cell. In our lattice approach an additional parameter is the size $L$ of the 1D lattice. The growth described in the previous section sets the time-unit by considering $v=1\, l/T$.

Following the seminal work of Axe and Yamada \cite{Axe} it is easy to realise that the model has a dimensional scaling. Using the nucleation rate $p$ (dimension $l^{-1} T^{-1}$) and the growth velocity $v$ (dimension $lT^{-1}$) we can define a natural length and time-scale for the KJMA growth process in 1D as
 \begin{eqnarray}
  \xi=(p/v)^{-1/2} \\
  \theta=(pv)^{-1/2}\,. 
 \label{scaling}
 \end{eqnarray}
This allows for any $G(s,t)$ function that is related with the space-time evolution of the system ($s$ is the spacial coordinate and $t$ the time-like coordinate) to be rescaled in a universal form by expressing $s$ and $t$ in the $\xi$ and $\theta$ units.

In the continuous growth model Kolmogorov \cite{Kolmogorov} derived a famous result 
\be
w(t)= \frac{W(t)}{L}=\exp[-V_{ex}(t)]
\ee{kolm-res}
for the fraction of untransformed material at time $t$. In the above equation $W(t)$ denotes the amount of material occupied by the growing cells, $L$ is the total size of the space where the growth takes place, and $V_{ex}(t)$ denotes the {\em total extended volume fraction}. The total extended volume is defined as the hypothetical volume of the cells at time moment $t$ if their growth would be unimpended by the other cells (i.e. in the absence of collision with other cells), divided by the size of the space where the growth takes place. The value of $V_{ex}(t)$ in 1D is
\be
V_{ex}(t)=\frac{1}{L}\int_0^t pL \, 2vt' \, dt'=pvt^2\,.
\ee{ext-vol}

Taking into account, that in our lattice model $v=1$, we get
\be
w(t)=\exp(-pt^2)\,.
\ee{KJMAw}
Following the works of Kolmogorov \cite{Kolmogorov} and Avrami \cite{Avrami} we can also estimate the number of nucleated cells $N(t)$ at time moment $t$ as
\be 
N(t)=Lp \int_0^t w(t') dt'=L\sqrt{p} \int_0^{t\sqrt{p}} \exp(-u^2)du\,.
\ee{KJMTNt}
For arbitrary $t$ time this leads to a non-analytical form, which can be expressed by using the {\em error function}. In the limit of $t\rightarrow \infty$ we get 
the total number of cells as
\be
N_t=N(\infty)=\frac{L}{2}\sqrt{\pi p}\,,
\ee{NtscalingKJMT}
which leads us to the average cell size $c$ in the final configuration:
\be
c=\frac{L}{N_t}=\frac{2}{\sqrt{\pi p}}\propto \frac{1}{\sqrt{p}}\,.
\ee{cKJMT} 
For our finite 1D lattice with size $L$ we can estimate also the time needed to get saturation. The time needed to get 
$w(t)=0$ is infinitely long. However, on our finite lattice the condition for saturation is to get the total number of non occupied sites
less than one. This leads to the saturation condition
\be
W(t_{sat})=1\,.
\ee{stopcond}
The estimate for the $t_{sat}$ time-length of the growth is
\be
t_{sat}=\sqrt{\frac{\ln{L}}{p}}\,.
\ee{tfinKJMT}

An exact theoretical method for getting the grain-size distribution in the 1D KJMA models is known for a quite long time, see for example the works of Axe and Yamada \cite{Axe}, Ben-Naim and Krapivsky \cite{Ben-naim} or Farjas and Roura \cite{Farjas}. The problem with these approaches is that it doesn't lead to a compact analytic form, only to a numerically computable density function, which is not practical for fitting experimental data. 

We consider thus an alternative, less accurate theory for the 1D KJMA growth, which yields an analytically compact form for the density function of the cell size-distribution. 
 
\section{Mean-field like approach}

Let us consider the 1D growth in a mean-field (MF) type approach. ``Mean-field'' means here that we assume no correlation effects between the growing cells, and we treat the distribution of empty sites as a completely random distribution.

We denote by $p$ the site activation probability and by $L$ the total number of lattice sites. At each time moment $t$ let $N(t)$ be the total number of Voronoi cells and $W(t)$ the empty (non activated) lattice sites. 

The probability that a randomly chosen site is empty at time moment $t$ is
\be
q(t)=\frac{W(t)}{L}\,.
\ee{e1}

The dynamics of the $N(t)$ and $W(t)$ quantities is described by the following equations:
\be
\frac{dN(t)}{dt}=pW(t)
\ee{dinN}
\be
\frac{dW}{dt}=-2q(t) N(t)-pW(t)\,.
\ee{dinW} 

The term $2q$ results in the following manner. Each Voronoi cell can growth in unit time step with probability $q^2$ on both of its side, leading to a 2 sites decrease in $W(t)$. The cell can growth in unit time with a probability $q(1-q)$ on one of its site. This site can be the left or the right one, both of them leading in $W(t)$ to a decrease with 1 site. In such a view the number of sites that are getting activated by the growth of the cell is $[2q^2+2q(1-q)]N(t)=2qN(t)$. The second term in equation (\ref{dinW}) results from the fact that each empty site can be activated with a probability $p$.

In such manner we get a solvable system of coupled first-order differential equations for $N(t)$ and $W(t)$. Using equation (\ref{e1}) we get
 \be
\frac{dN(t)}{dt}=pW(t)
\ee{dinNf}
\be
\frac{dW}{dt}=-2\frac{W(t)}{L} N(t)-pW(t)\,,
\ee{dinWf} 
 which should be solved with the initial conditions
 \be
 N(0)=0
 \ee{initN}
 \be
 W(0)=L\,.
 \ee{initW}
 
The solution of the system is
\be 
N(t)=-\frac{1}{2}pL+\frac{1}{2}L\sqrt{p(p+4)}F(t)
\ee{solN}
\be
W(t)=L\left(1+\frac{p}{4}-F(t)^2-\frac{p}{4}F(t)^2\right)\,,
\ee{solW}
where
\be
F(t)=\tanh{\left[ \frac{1}{2}\left( \sqrt{p(4+p)}t+2\tanh^{-1}{\sqrt{\frac{p}{4+p}}}\right)\right]}\,.
\ee{Ft}

In the limit of $p \ll1$, which is our case, we can keep only the leading terms in $p$ and we get the simplified solution of the system
\be 
N(t)\approx L\sqrt{p}F(t)
\ee{solNsim}
\be
W(t)\approx L(1-F(t)^2)\,,
\ee{solWsim}
where
\be
F(t)\approx \tanh{\left[\sqrt{p}(t+\frac{1}{2})\right]}\,.
\ee{Ftsim}

Since one time unit is equivalent with one step in the growth process, apart of the very beginning of the dynamics we can assume
$(t+1)\approx t$, leading to
\be
F(t)\approx \tanh{\left(\sqrt{p}t\right)}\,.
\ee{Ft}

According to these results, the {\bf total number of cells}, $N_t$ at the end of the growth process can be approximated as:
\be 
N_t=\lim_{t\rightarrow \infty} N(t)\approx L\sqrt{p}\,.
\ee{totalcellnum}
The {\bf mean cell size} $c$ will be given by:
\be
c=\lim_{t\rightarrow \infty} \frac{L}{N(t)}\approx \frac{1}{\sqrt{p}}\,.
\ee{meancellsize}
The scaling property as a function of $p$ is the same as the result (\ref{cKJMT})
 given by the KJMA theory. The saturation time necessary to fill up all sites $t_{sat}$, will be  estimated now in the same manner as in the classical KJMA theory
\be
W(t_{sat})=1\,,
\ee{longtime}
and consequently
\be
F(t_{sat})=\sqrt{1-\frac{1}{L}}\,.
\ee{eqFtlong}
From here one gets
\be
t_{sat}=\frac{1}{\sqrt{p}} \tanh^{-1} {\sqrt{1-\frac{1}{L}}}\,.
\ee{solt2}

Taking into account that $1/L=\epsilon \ll1$, we can perform a Taylor expansion of the $\tanh^{-1}(\sqrt{1-\epsilon})$ term around 
$\epsilon=0$, and we get
\be
\tanh^{-1}(\sqrt{1-\epsilon})\approx \ln(2)-\frac{1}{2}\ln(\epsilon)-\frac{1}{8}\epsilon+O(x)^2\,.
\ee{taylor}
This leads us to
\be
 t_{sat} \approx \frac{1}{2\sqrt{p}}\ln(4L)
\ee{timefinal}
suggesting for constant $L$ the same scaling as the KJMA theory in 1D (\ref{tfinKJMT})
\be
t_{sat} \propto \frac{1}{\sqrt{p}}\,.
\ee{timescaling} 
The finite-size effects, i.e. variation of $t_{sat}$ as a function of $L$, is however different. 

We proceed now to determine the {\bf size-distribution function} of the cells in the limit where we neglect $q(t)^2$ and keep only the leading terms in $p$, i.e. where the solutions given by equations (\ref{solNsim}), (\ref{solWsim}), (\ref{Ftsim}) are valid. 

If we denote the number of cells of size $k$ at time moment $t$ by $N(k,t)$, the dynamics of the system can be written by a coupled system of master equations. For cells of sizes 1 and 2, the equations are a little different, but for $k>2$ the growth equations have the same form:
\small
\begin{align}
& \frac{dN(1,t)}{dt}=pLq-N(1,t)(2q-q^2)  \nonumber \\
& \frac{dN(2,t)}{dt}=N(1,t)2q(1-q)-N(2,t)(2q-q^2) \nonumber \\
&...  \nonumber\\
& \frac{dN(k,t)}{dt}=N(k-2,t)q^2+N(k-1,t)2q(1-q)- \nonumber\\
&- N(k,t)(2q-q^2)  \nonumber \\
&...
\label{Nmaster}
\end{align}
\normalsize

Keeping only the first order terms in $q$:
\begin{align}
& \frac{dN(1,t)}{dt}=pLq-2qN(1,t)  \nonumber \\
&...  \nonumber\\
& \frac{dN(k,t)}{dt}=2qN(k-1,t)-2q N(k,t)  \nonumber \\
&...
\label{Nmasterq}
\end{align}

We introduce the 
\be
P_i(k,t)=\frac{N_i(k,t)}{N(t)}
\ee{probabdef}
probabilities of finding a cell with size $k$ among all cells in the system at time moment $t$.
It's first order time derivative is:
\be 
\frac{dP_i(k,t)}{dt}=\frac{1}{N(t)}\frac{dN(k,t)}{dt}-\frac{1}{N(t)^2}\frac{dN(t)}{dt}N(k,t)\,.
\ee{NtoP}

Using equations (\ref{e1}) and (\ref{dinN}) we can write:
\be
\frac{dP_i(k,t)}{dt}=\frac{1}{N(t)}\frac{dN(k,t)}{dt}-pq(t)P(k,t)\frac{L}{N(t)}\,.
\ee{NtoPfin}

 We rewrite now the master equations (\ref{Nmasterq}) for the corresponding probabilities:
 \begin{align}
& \frac{dP(1,t)}{dt}=\frac{L}{N(t)}pq-2qP(1,t)-pqP(1,t)\frac{L}{N(t)} \nonumber \\
&...  \nonumber\\
& \frac{dP(k,t)}{dt}=2qP(k-1,t)q- 2qP(k,t)- pqP(k,t)\frac{L}{N(t)}  \nonumber \\
&...
\label{Pmasterq}
\end{align}
The master equation for the cumulative distribution function 
\be
S(k,t)=\sum_{i=1}^{k} P(k,t)\,.
\ee{cumdist}
can be obtained by adding up the equations in (\ref{Pmasterq}):
\be
\frac{\partial S(k,t)}{\partial t}=-[S(k,t)-S(k-1,t)]2q+\frac{pqL}{N(t)} [1-S(k,t)]\,.
\ee{cume1}

Let us consider now the continuous limit of this probability distribution, and instead of $P(k,t)$ let us use the probability density $\Omega(s,t)$, where $s$ is a continuous variable. Equation (\ref{cume1}) becomes now a partial differential equation (PDE) of the form
\be
\frac{\partial \Omega(s,t)}{\partial t}=-2q\frac{\partial \Omega(s,t)}{\partial s}+\frac{pqL}{N(t)} \left[1-\Omega(s,t)\right]\,.
\ee{cont}

Using the result (\ref{solNsim}) for $N(t)$, and (\ref{solWsim}) for $W(t)$, we get the following PDE for the distribution function:
\be
\frac{\partial \Omega(s,t)}{\partial t}\frac{F(t)}{1-F(t)^2}+2F(t)\frac{\partial \Omega(s,t)}{\partial s}=  \sqrt{p} \left[1-\Omega(x,t)\right]\,.
\ee{pde}

For $F(t)$ given by equation (\ref{Ft}) one obtains
\begin{align}
&\frac{\partial \Omega(s,t)}{\partial t}\frac{\tanh(t\sqrt{p})}{1-\tanh^2(t\sqrt{p})}+\frac{\partial \Omega(s,t)}{\partial s}\tanh(t\sqrt{p})=  \nonumber \\
&= \sqrt{p} \left[1-\Omega(s,t)\right]\,.
\label{pdew}
\end{align}

This is an equation that independent of $L$. We can also show that the evolution equation for the cumulative distribution function will be independent of the parameter $p$ if we use the cell size relative to the mean cell size and rescale the time properly. More precisely, let us consider the following scalings: $x=s/c=s\sqrt{p}$ (where $c$ denotes the average cell size) and $\tau=t\sqrt{p}$. With these new variables we get
\begin{align}
&\frac{\partial\Omega(x,\tau)}{\partial\tau}\frac{\tanh(\tau)}{1-\tanh^2(\tau)}+2\frac{\partial\Omega(x,\tau)}{\partial x}\tanh(\tau)=  \nonumber \\
&= \left[1-\Omega(x,\tau)\right]\,,
\label{pdew}
\end{align}
which is obviously independent of $p$, suggesting a scaling property for the cell-size distribution function. This result confirms again the predicted scaling of the mean cell size (\ref{meancellsize}) and time needed for saturation (\ref{timescaling}) as a function of $p$.

The general solution of this first order partial differential equation obtained with the standard mathematical methods writes as
\be
\Omega(x,\tau)=1-\frac{1}{\tanh{(\tau)}} e^{H[x-2\tanh{(\tau)}]}\,,
\ee{sol}
with $H[z]$ an arbitrary function. 

Since $\Omega(x,\tau)$ is the cumulative cell size distribution function, we search for a general solution satisfying the criteria $\Omega(\infty,\tau)=1$, $\Omega(0,\infty)=0$ and $d\Omega(x,\tau)/dx>0$ for all $x$ values. Taking into account that $0\le\tanh(x)\le1$ a general class of function $H[z]$ satisfying the imposed conditions is $H[z]=- \gamma (z+2)^\alpha$, where $\gamma$ and $\alpha$ are two positive  constants. This leads to the cumulative cell size distribution function
\be
\Omega(x,\tau)=1-\frac{1}{\tanh{(\tau)}} e^{-\gamma (x-2\tanh{(\tau)+2)^\alpha}}\,.
\ee{solfin}
It is easy to verify that such a solution satisfies the partial differential equation (\ref{pdew}). The above solution leads for $\tau\rightarrow \infty$ the final cumulative cell size distribution function:
\be
\Omega(x)=1- e^{-\gamma x^\alpha}\,.
\ee{solstat}

 The corresponding probability density function is
 \be
 \rho(x)=\gamma \alpha x^{\alpha-1}e^{-\gamma x^\alpha}\,,
 \ee{weibtheo}
which is the well-known Weibull distribution. Taking into account that $\langle x \rangle=1$, we get
\be
\gamma=\left[\Gamma\left(1+\frac{1}{\alpha}\right)\right]^\alpha\,,
\ee{rel}
with $\Gamma(x)$, the classical Gamma function.

\section{Computer simulations}

In order to check the theoretical predictions Monte Carlo type computer simulations have been performed. We have considered system sizes up to $L_{max}=10^9$ lattice sites and varied the nucleation probability in the range of $10^{-2}\le p \le 10^{-7}$.

\begin{figure}[h]
\centerline{\includegraphics[width=0.45\textwidth]{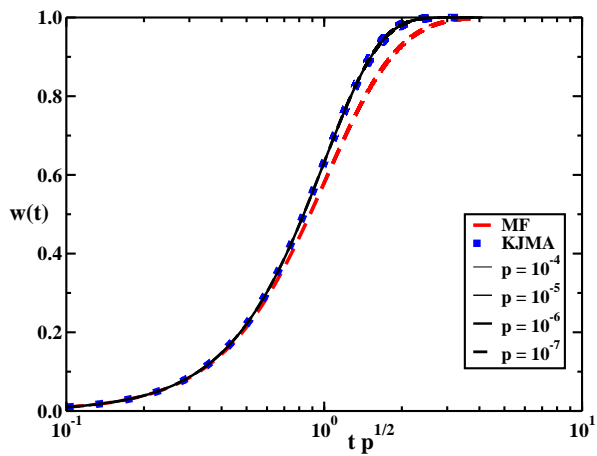} }
\caption{~\label{timeevol}
Fraction of activated sites (transformed phase), $w=W(t)/L$, as a function of time. Continuous curves are computer simulation results obtained for lattices with $L=10^8$ sites and for various nucleation probabilities $p$, averaged on $Q=100$ realisations. Blue squares are the prediction of the KJMA theory, while the red dashed line is the result given by the MF theory. } 
\end{figure}

First, we have studied the statistical properties for the time-evolution of the system. Considering lattices with $L=10^8$ sites and various nucleation probability values we followed the fraction of transformed  phase $w(t)=W(t)/L$ as a function of time. Simulation results averaged on $Q=100$ realisations are plotted with continuous lines on Figure \ref{timeevol}. The prediction of the KJMA theory (blue squares) describes well the simulation results. As expected, our MF theory (dashed red line) gives a fair description at the beginning of the dynamics, where the cell's growth can be considered independent, and a considerable deviation is observed for later time-steps. 

\begin{figure}[h]
\centerline{\includegraphics[width=0.45\textwidth]{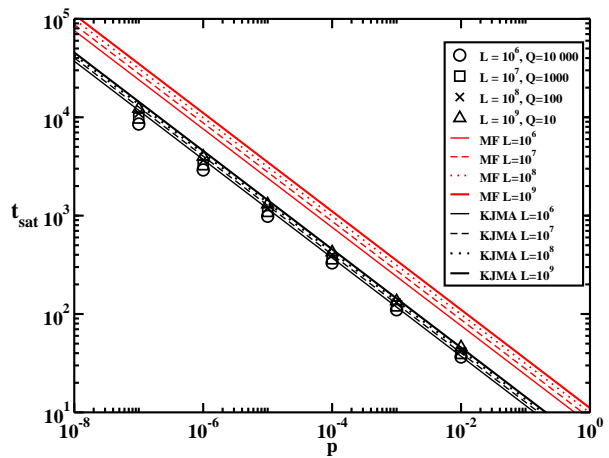} }
\caption{~\label{tfin-pscale}
Time needed for saturation, $t_{fin}$ as a function of the nucleation probability, $p$. Computer simulation results (circles) are obtained on lattice 
sizes $L$ and averaged on $Q$ realizations as indicated in the legend. Black lines represent the prediction (\ref{tfinKJMT}) of KJMA theory. Red lines indicate our MF prediction given by equation (\ref{timefinal}). } 
\end{figure}

\begin{figure}[htb]
\centerline{\includegraphics[width=0.45\textwidth]{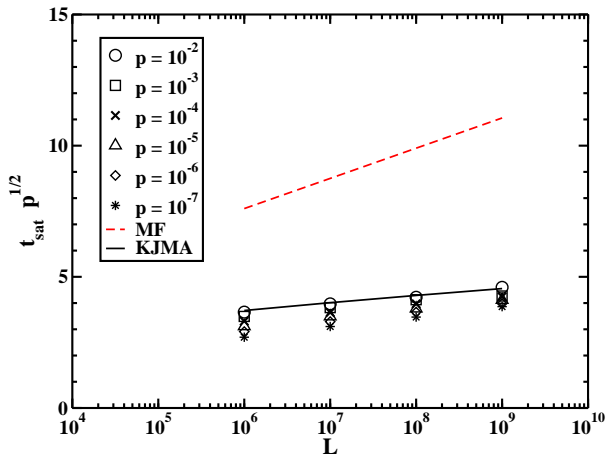} }
\caption{~\label{tfin-Lscale}
Saturation time, $t_{fin}$ as a function of the system size. Simulation results (symbols) are for different $p$ values indicated in the legend. The data is averaged on many configurations $Q$ which is changing with the system size $L$ (the same $Q$ values are used as in case of Figure \ref{tfin-pscale}).  The dashed line indicates the upper bond given by equation (\ref{timefinal}). } 
\end{figure}

The $t_{sat}$ time-length of the growth process until saturation will be in our focus now. The results are presented on Figures \ref{tfin-pscale} and \ref{tfin-Lscale}. On Figure \ref{tfin-pscale} symbols represents the computer simulation results obtained for the scaling of the saturation time as a function of the nucleation probability. Results for systems with different sizes $L$ are presented  as it is indicated in the legend. The results are averaged on many configurations, indicated by the $Q$ values in the legend. Black lines indicate the prediction obtained from the KJMA theory (\ref{tfinKJMT}) and red lines indicate our MF prediction given by equation (\ref{timefinal}). The observed trend suggest that the scaling predicted by both theories, $t_{sat}\propto p^{-1/2}$ is correct. The actual values for $t_{sat}$ given by the KJMA theory are however much better in this case than the results of the MF theory. 

The saturation time depends as a function of the system size $L$, as it is visible in the computer simulation results plotted in Figure \ref{tfin-pscale} and in agreement with the prediction of the theoretical attempts. This finite-size effect is studied for different nucleation probabilities and the results are plotted on Figure \ref{tfin-Lscale}. In order to collapse the data for different $p$ values we have plotted $t_{sat}p^{1/2}$ as a function of the system size. Symbols represent computer simulation data for different $p$ values. The data is averaged on many configurations $Q$ that is changing with the system size $L$. Here, the same $Q$ values are used as in case of Figure \ref{tfin-pscale}. The continuous black line indicates the KJMA prediction (\ref{tfinKJMT}), while the red dashed line is the result (\ref{timefinal}) given by the MF theory. The result (\ref{tfinKJMT}) obtained from the KJMA theory gives a good approximation for the computer simulation data. The increasing trend of the MF theory is correct, however the actual results offer a much weaker approximation than the ones of the KJMA theory. One should also note, that the result of the adapted KJMA theory is also not perfect. The data for different $p$ values suggests that even the $t_{sat}\propto p^{1/2}$ scaling hypothesis is not rigorously exact! This difference is suggested also by the slightly different slopes in Figure \ref{tfin-pscale}.

\begin{figure}[h!]
\centerline{\includegraphics[width=0.45\textwidth]{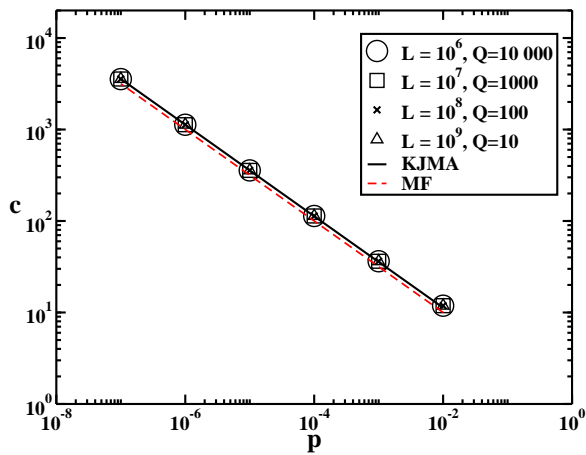} }
\caption{~\label{cscale}
Mean cell size as a function of the $p$ nucleation probability. Computer simulations were realised on lattices with different sizes $L$ and averaged on $Q$ realizations as it is indicated in the legend. The continuous black line represents the prediction of the KJMA theory (\ref{cKJMT}) and the red dashed line is our MF prediction (\ref{meancellsize}). }
\end{figure}

We discuss now the statistical results for the final cell-size distribution. Results for the mean cell size, $c$, as a function of the nucleation probability are given in Figure \ref{cscale}. Here, the symbols represent simulation results for different $L$ system sizes, and averaged on $Q$ realisations (as indicated in the legend). The continuous black line indicates the result of the KJMA theory (\ref{cKJMT}) and with red dashed line we plot our MF prediction given in equation (\ref{meancellsize}). The computer experiments confirm nicely the $N_t\propto p^{-1/2}$ scaling. Although the KJMA theory offers a perfect description for the results, the values predicted in the MF approach (\ref{meancellsize}) gives a surprisingly good approximation. 

\begin{figure}[h]
\centerline{\includegraphics[width=0.45\textwidth]{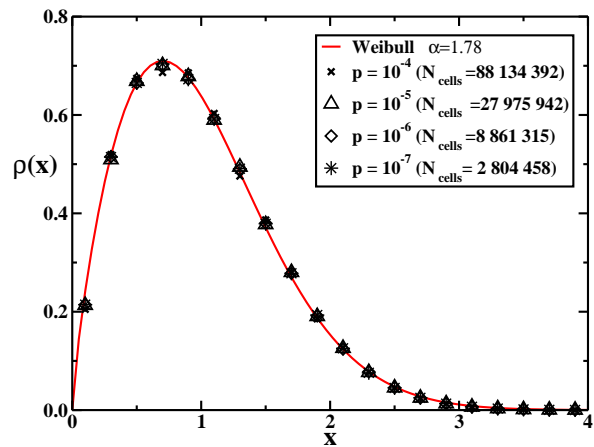} }
\caption{~\label{weibull}
Final cell size distributions, $\rho(x)$, obtained for different $p$ nucleation probabilities (as indicated in the legend) and $L=10^8$ system size. The continuous curve shows a Weibull fit with $\alpha=1.78$ for the collapsed data. Simulation results are generated from $Q=100$ different runs, which result in $N_{cells}$ individual cells (indicated in the legend). } 
\end{figure}

The final cell-size distribution function was studied for different system sizes and nucleation probabilities. For the $\rho(x)$ probability density functions, where the cell size is normalised by the mean cell-size ($x=s/c$), the collapse of the distribution functions for different $p$ and $L=10^8$ sites are shown on Figure \ref{weibull}. Here, the distribution function is obtained from $Q = 100$ different simulations resulting in $N_{cells}$ individual cells indicated in the legend. On the same Figure we also indicate that a Weibull fit (see equation \ref{weibtheo}) with $\alpha=1.78$
works well. 

\begin{figure}[h]
\centerline{\includegraphics[width=0.45\textwidth]{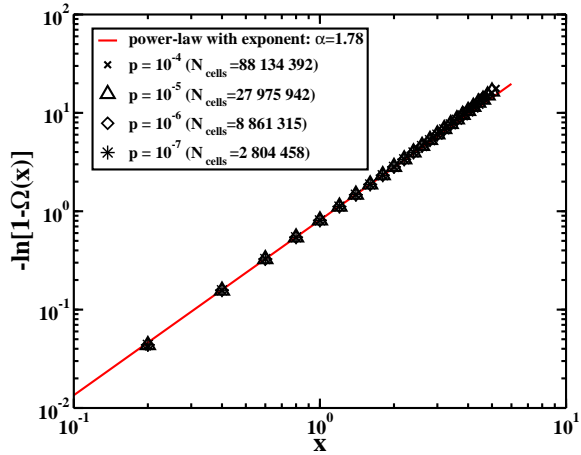} }
\caption{~\label{weibull-cum}
Scaling properties related to the final cumulative cell-size distribution. The linear trend on the log-log plot for $-\ln[1-\Omega(x)]$ suggests that the Weibull distribution function with $\alpha=1.78$ describes well the computer simulation data. Symbols indicate computer simulation results for different nucleation probabilities, as it is specified in the legend. The continuous red line is a power-law with exponent $1.78$. Simulation results were obtained on lattices with $L=10^8$ sites and calculated from $Q=100$ different runs which result in $N_{cells}$ individual cells (indicated in the legend). } 
\end{figure}

To further argument the Weibull form of the final cell-size distribution function, we have plotted $-\ln[1-\Omega(x)]$ as a function of $x$ on Figure \ref{weibull-cum} (we remaind that $\Omega(x)$ is the cumulative distribution function). If one accepts for $\Omega(x)$ the Weibull distribution given by equation (\ref{solstat}), than a power law trend is expected for $-\ln[1-\Omega(x)]$. The straight trend of the simulation results on a log-log plot indicates a fair scaling with an exponent of $\alpha=1.78$. This gives us further evidence, that the Weibull distribution is appropriate for describing the final cell-size distribution for the 1D KJMA lattice model with continuous nucleation.

\section{Conclusions}

We have considered a KJMA growth process with continuous nucleation on one-dimensional lattices. In order to investigate the growth dynamics, the total time of the growth process and the statistics of the final cell-size distribution we used both the classical KJMA theory and a mean-field (MF) type approximation. Computer simulation results were compared with the predictions of these theories. We found that the KJMA theory offers an excellent description for the statistical properties of the growth process, however the lack of a compact form for the cell-size distribution function is a great impediment. The MF type approximation gives a good description for the initial part of the dynamics, where the growth of the cells can be considered as independent and coalescence is not important. The MF theory leads to the same scaling properties for the saturation time and mean-cell-size as a function of the nucleation probability as the KJMA theory. However, the advantage of the MF type approach is that it leads to a compact analytical approximation for the final cell-size distribution function. According to this, we expect a Weibull-type distribution. Computer simulations confirm that the Weibull distribution is a proper fit for the final cell-size distribution. This result can be of importance in many practical applications for fitting the experimental data.




\end{document}